\documentclass[twocolumn,tightenlines,showpacs,aps]{revtex4}
\usepackage{graphicx}

\begin{document}

\title{A conclusive demonstration of vibrational pumping and determination of SERS cross sections}

\author{R. C. Maher$^1$} \email{robert.maher@imperial.ac.uk}

\author{P. G. Etchegoin$^2$} \email{Pablo.Etchegoin@vuw.ac.nz}

\author{E. C. Le Ru$^2$} \email{Eric.LeRu@vuw.ac.nz}

\author{L. F. Cohen$^1$}

\affiliation{$^1$The Blackett Laboratory, Imperial College London\\
Prince Consort Road, London SW7 2BZ, United Kingdom}

\affiliation{$^2$The MacDiarmid Institute for Advanced Materials and Nanotechnology\\
School of Chemical and Physical Sciences\\ Victoria University of Wellington\\
PO Box 600, Wellington, New Zealand}

\date{\today}

\begin{abstract}
We provide  conclusive demonstration of vibrational pumping under
Surface Enhanced Raman Scattering (SERS) conditions by performing
anti-Stokes/Stokes ratio measurements down to 10\,K using dried
silver colloids, the dye rhodamine 6G and 676\,nm laser
excitation. The method we propose allows the measurement of the
SERS cross sections for different modes and the determination of
the asymmetry between the anti-Stokes and Stokes SERS cross
sections.
\end{abstract}


\maketitle

After its discovery in the 1970's, Surface Enhanced Raman
Scattering (SERS) became a subject of intense research up to the
mid-1980's with varied applications\cite{Moskovits85}. A
resurgence in interest occurred after two independent reports
\cite{Nie97, Kneipp97} in 1997 on the observation of single
molecule SERS, with potential advantages over fluorescence-based
techniques. In addition, the possibility of observing vibrational
pumping via SERS was proposed in 1996\cite{pumping}. This proposal
has been much debated for almost a decade in the
literature\cite{UsFaraday} but a clear-cut confirmation has
remained elusive. Its importance lies not only in fundamental
aspects of molecular spectroscopy but also in the potential of the
technique for metrology applications and the determination of SERS
cross sections\cite{pumping}.

In this letter we provide conclusive evidence for vibrational
pumping in SERS. A brief historical account of the subject is
beneficial at this stage for the forthcoming discussion.

Consider a single vibrational level in a molecule at temperature
$T$. A laser with intensity $I_L$ [W/cm$^2$] creates occupation of
the level through pumping with a rate proportional to both the
Raman-Stokes cross section $\sigma_S$ [cm$^2$] and the intensity.
The population of the level is in part also determined by the
thermal equilibrium condition defined by the molecular temperature
$T$. Vibrations remain in a specific level for a finite lifetime
$\tau$ [s], which takes into account all the possible relaxation
mechanisms starting from that level, either through intramolecular
vibrational relaxation IVR (anharmonic processes) or external
relaxation mechanisms. The rate equation for the average phonon
population $n$ of the vibrational mode is:
\begin{equation}
\frac{dn}{dt}=\frac{\sigma_S
I_L}{\hbar\omega_L}+\frac{n_B(\hbar\omega_v/k_B
T)}{\tau}-\frac{n}{\tau},\label{rate}
\end{equation}
where $\hbar\omega_L$ and $\hbar\omega_v$ [J] are the energies of
the laser and vibration, respectively, and $n_B(\hbar\omega_v/k_B
T)$ is the Boltzmann factor for that vibration. The second term
corresponds to thermal excitation and has been adjusted so that in
the steady state ($dn/dt=0$) $n=n_B(\hbar\omega_v/k_B T)$ when
$I_L=0$. When $I_L \neq 0$:
\begin{equation}
n=\frac{\tau \sigma_S I_L}{\hbar\omega_L}+e^{-\hbar\omega_v/k_B
T}\label{n2},
\end{equation}
where the first and second terms account for vibrational pumping
and thermal effects, respectively. This expression can be
justified more rigorously\cite{UsFaraday}. This is the weak
pumping regime where vibrational pumping might compete with the
thermal population of the level but $n$ remains small. The Stokes
Raman signal is independent of $n$ and it is simply $I_S=N
\sigma_S I_L$, where $N$ is the number of molecules. In contrast,
the anti-Stokes signal does depend on $n$ and it is given by:
$I_{aS}= n N \sigma_{aS} I_L$. The anti-Stokes/Stokes ratio
$(\rho)$ of a specific vibration is then:

\begin{equation}
\rho= I_{aS}/I_S= \frac{\sigma_{aS}}{\sigma_S}~n= A
\left[\frac{\tau \sigma_S
I_L}{\hbar\omega_L}+e^{-\hbar\omega_v/k_B T}\right], \label{rho}
\end{equation}

where $A\equiv \sigma_{aS}/\sigma_S$ is the {\it asymmetry
parameter} taking into account that in SERS we have
$\sigma_{aS}\neq \sigma_S$ in general, due to the presence of
underlying frequency-dependent plasmon resonances\cite{RobJCP}. If
the first term (pumping) in Eq. (\ref{n2}) dominates over the
second (Boltzmann factor), this scenario produces a Stokes signal
$\propto I_L$, an anti-Stokes one $\propto I_L^2$, whilst $\rho$
is linear in $I_L$. Moreover, the slope of $\rho$ vs. $I_{L}$
provides $A\tau\sigma_S/\hbar\omega_L$. If $A$ is known or
estimated, and if a judicious estimation of $\tau$ can be made,
the SERS Stokes cross section can then be gained. This description
led to a series of claims on the order of magnitude of the SERS
cross sections\cite{pumping}. The quadratic dependence of $I_{aS}$
on $I_L$ at room temperature was the origin of the original claim
on SERS pumping\cite{pumping}.

The evidence for SERS pumping has been much resisted in the
literature\cite{Haslett}. There have been entire publications
dedicated to reproducing these experiments under similar
conditions without successfully confirming the
results\cite{Haslett,Brolo}. The picture has remained
inconclusive, accordingly, and has been attributed to a mixture of
causes from sample variability to underlying plasmon resonances
and laser heating. For example the power dependence of $\rho$ can
be explained purely by heating effects\cite{Brolo}. If the first
term is negligible compared to the second one in Eq. (\ref{rho})
(normal situation) then $\rho\propto e^{-\hbar\omega_v/k_B T}$. If
the laser causes local heating, the molecule is at a higher
temperature $T^{*}=T+\Delta T\equiv T+ \beta I_L$. When $\Delta T
\ll T$, expanding this expression, we obtain $\rho\propto {\rm
exp}(\hbar\omega_v \beta I_L/ k_B T^2)$; i.e. an exponential or
linear dependence on $I_L$ (depending on the value of
$\hbar\omega_v \beta/ k_B T^2$) can be expected. These
dependencies have indeed been observed\cite{UsaSSratios}, and they
can easily be confused with pumping.

The solution to this problem comes from SERS measurements at
temperatures much lower than any reported
previously\cite{RobFaraday,Fukayama}. The quantity $\tau \sigma_S
I_L/\hbar\omega_L$ is normally $\ll 1$ and it is completely
negligible under normal Raman scattering conditions. Under SERS
conditions, this quantity is still small but can become comparable
to ${\rm exp}(-\hbar\omega_v/k_B T)$ if $I_L$ is sufficiently
high\cite{pumping}, in which case heating effects can complicate
the interpretation, or if $T$ is sufficiently small. There is then
a crossover from a anti-Stokes/Stokes ratio dominated by the
thermal population $(\rho \propto {\rm exp}(-\hbar\omega_v/k_B
T))$ to one dominated by pumping $(\rho\propto \tau \sigma_S
I_L/\hbar\omega_L)$. This crossover is easily observed in the
following experiments at temperatures in the range $\sim 10-150$ K
depending on the analytes and the specific mode under
consideration. The temperature at which the crossover happens
depends on mode energy due to the rapid variation of the
exponential term. This will occur at higher temperatures the
larger the Raman shift of the mode, with small variations from one
case to another depending on the magnitude of the first term in
Eq. (\ref{rho}), which depends on the specific $\sigma_S$ and
$\tau$.

The reason why vibrational pumping is very hard to identify at
higher temperatures is because it involves the study of small
departures in mode populations from an already existing thermal
population in the levels. By going to very low temperatures,
departures from zero population in the levels become more evident.
Once the crossover for different modes has been reached, it is
possible to observe anti-Stokes signals which are not related at
all to a thermal population and only depend on the relative
$\sigma$'s and $\tau$'s. At sufficiently low temperatures, all
main Raman peaks become independent of $T$ on the anti-Stokes side
and have relative intensities which are not related to a Boltzmann
factor. For example, at 10 K, we can observe modes with
$\hbar\omega_v\sim 1500-1600$ cm$^{-1}$ with a measurable
intensity in a few seconds on the anti-Stokes side (the Boltzmann
factor for these modes at 10 K is $\sim 10^{-100}$) while modes
with smaller energies are much weaker in intensity or, in some
cases, barely visible. The breakdown of relative intensities and
the non-Boltzmann distribution of signals on the anti-Stokes side
are both confirmation of SERS pumping. It can also be said that
each mode has an effective temperature $T_{eff}$ which is
different for each mode, thus showing explicitly the breakdown of
internal thermal equilibrium of the molecule.

We concentrate now on the experimental evidence for the much
studied dye rhodamine 6G (RH6G). Silver colloids were produced as
described in Ref.\cite{Miesel}. SERS samples were prepared to a
final 1 $\mu$M concentration of RH6G in a 50\%-50\% solution of
colloids and 20 mM KCl. The sample was then dried onto a Si
substrate and mounted on a closed-cycle He-cryostat
(CTI-cryogenics) with temperature control in the range 10-300 K.
Raman measurements have been performed for several laser lines of
a Kr$^{+}$ and Ar$^{+}$-ion lasers; we shall concentrate here only
on a subset of these measurements to prove the point we raised.
The laser was focused to a 20 $\mu$m spot in diameter and the
signal was collected by a high-numerical aperture photographic
zoom lens (Canon, $\times 10$ magnification) through the quartz
window of the cryostat and onto the entrance slit of a
high-dispersion double-additive U1000 Jobin-Yvon spectrometer
coupled to a N$_2$-cooled CCD detector. The wider spot area
compared to those used in Raman microscopy has the triple
advantage of: $(i)$ reducing photobleaching to a negligible level;
$(ii)$ reducing any indirect laser heating effect; and $(iii)$
improving the averaging over clusters geometries.

\begin{figure}
\centering{
 \includegraphics[width=8cm]{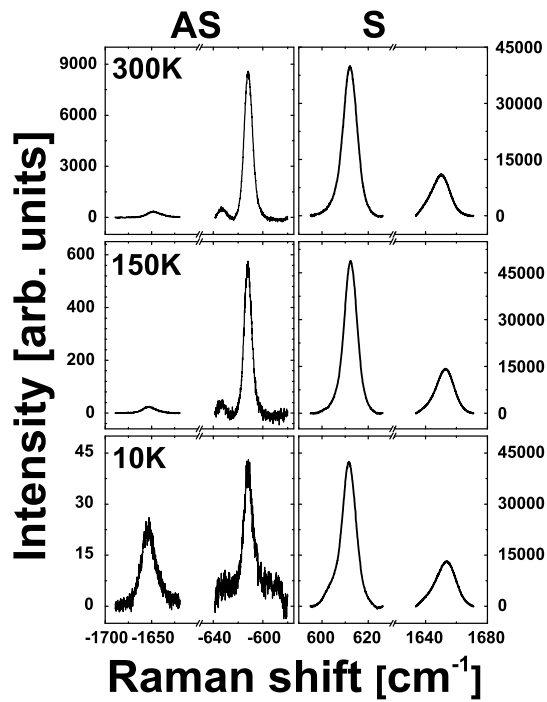}
} \caption{ Anti-Stokes (left) and Stokes (right) spectra for two
widely separated modes of RH6G (610 cm$^{-1}$ and 1650 cm$^{-1}$)
under 676 nm laser excitation. At room $T$ the normal ratio is
observed (slightly modified by the asymmetry parameter $A$). At
150 K the anti-Stokes intensity of the 1650 cm$^{-1}$ mode is
already anomalous with respect to the 610 cm$^{-1}$. At 10 K the
anti-Stokes signals of both peaks are comparable in size and only
existent because of vibrational pumping. This is equivalent to
having both modes with different effective temperatures (Table
\ref{tabla}).}
 \label{fig1}
\end{figure}

Figure \ref{fig1} shows anti-Stokes/Stokes spectra for two strong
Raman modes widely separated in energy (610 and 1650 cm$^{-1}$,
respectively) for 676 nm laser excitation (Kr$^{+}$-ion laser,
$I_L=1.6~10^{8}$ W/m$^2$) at 3 different temperatures. At 300 K
the peaks show the normal anti-Stokes/Stokes ratio (except for the
asymmetry factor $A$ which is a small correction compared to the
effect of the exponential in (\ref{rho})). The Stokes side remains relatively constant throughout the measurement as can be seen in the figure. At $T=$150 K an anomalous relative intensity between the two peaks
on the anti-Stokes side can already be seen while the anti-Stokes
signal of the 610 cm$^{-1}$ mode has decreased (as expected by the
lower temperature) the one for the 1650 cm$^{-1}$ mode still
remains visible. This is because, at this $T$, the pumping term
already dominates (1$^{st}$ term in Eq. (\ref{rho})) while the
former is still responding to temperature changes (2$^{nd}$ term
in Eq. (\ref{rho})), as we shall show in the next figure. At
$T=10$ K, the anti-Stokes signals of both peaks are almost
comparable in size. The Boltzmann factors for these two peaks at
10 K are $\sim 10^{-38}$ (610 cm$^{-1}$) and $\sim 10^{-100}$
(1650 cm$^{-1}$); i.e. the only reason why they are still seen is
SERS pumping.

\begin{figure}
\centering{
 \includegraphics[width=8cm]{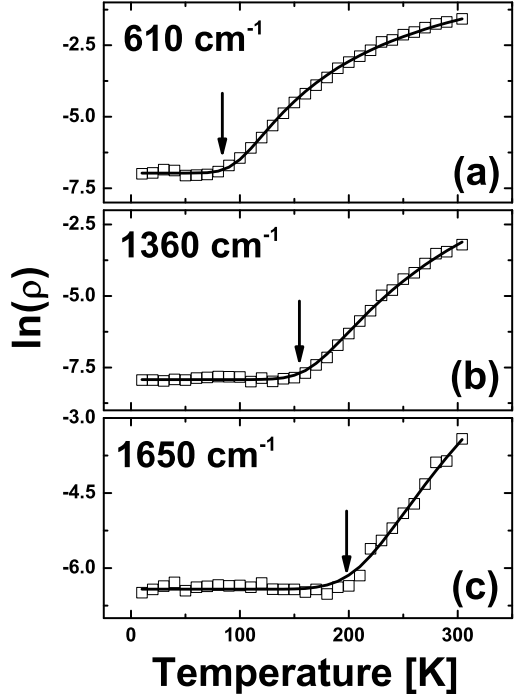}
} \caption{Anti-Stokes/Stokes ratio (ln$(\rho)$) as a function of
$T$ for three modes in RH6G (676 nm laser excitation). Note that
the crossover happens at different temperatures for different
modes (vertical arrows), being at higher $T$'s for larger
vibrational energies. The crossover temperature is roughly
equivalent to the effective mode temperature $T_{eff}$. The solid
lines are fits using Eq. (\ref{fit}). From the two parameters of
the fit, the FWHM of the peaks, and the laser intensity $I_L$, the
SERS cross section can be readily estimated.}
 \label{fig2}
\end{figure}

Figure \ref{fig2} shows the experimental anti-Stokes/Stokes ratios
$({\rm ln}(\rho))$  for 3 different modes in RH6G as a function of
temperature for 676 nm laser excitation. A clear crossover between
the two regimes dominated by the first or second term in Eq.
(\ref{rho}) can be seen in the data. The crossover happens at
higher temperatures for higher energy modes, as expected. It is
convenient to fit the experimental data to:

\begin{equation}
{\rm ln}(\rho)=a+ {\rm ln}\left[b+e^{-\hbar\omega_v/k_B T}\right],
\label{fit}
\end{equation}
where $a=$ln$(A)$, and $b=\tau \sigma_S I_L/\hbar\omega_L$. This
is then a fit with two parameters, from where $\tau \sigma_S$ can
be gained for a known $I_L$. Fits to Eq. (\ref{fit}) are
explicitly shown in Fig. \ref{fig2}. The agreement between theory
and experiment is excellent. From here we can obtain $\tau
\sigma_{S}$ and eventually $\sigma_{S}$ if the lifetime is
estimated.

The need to obtain the relaxation time $\tau$ for each level might
be seen as a drawback at first, but it is not a fundamental one.
For a start, relaxation times could be ultimately determined by
invoking a different type of spectroscopy (time resolved) if
needed, but even within the present framework they can also be
estimated from the FWHM $(\Gamma)$ of the peaks. The plateaus in
Fig. \ref{fig2} at low temperatures suggest that $\tau$ remains
constant (within a few \%) in this temperature range, in
accordance with the almost constant $\Gamma$ observed. If the
broadening is homogeneous\cite{book}, $\tau$ follows directly from
$\tau\sim \hbar/\Gamma$. Molecular vibrations have such strong
couplings with other vibrations in the molecule that contributions
from inhomogeneous broadenings are negligible. Even in SERS
experiments where the single molecule limit is
approached\cite{UsJPC} (which would be more sensitive to
inhomogeneous broadening) the FWHM does not change by more than
$1-1.5$ cm$^{-1}$ in peaks with a typical $\Gamma$ of $\sim 15-20$
cm$^{-1}$. An estimation of $\tau$ from $\Gamma$ is expected to be
accurate within $\sim 10-15$ \%, accordingly. An almost constant
$\tau$ at low temperatures is also expected from general
considerations on anharmonic interactions where $\Gamma$ is
expected to follow $\Gamma= \Gamma_0 (2 n_{{\rm
Bose}}(\hbar\omega_{av}/k_B T))+1)$, with $\hbar\omega_{av}$ being
a characteristic energy in the density of states of the molecule.
For typical molecular vibrations in organic dyes this can be
neglected at low $T$'s. By taking these considerations into
account, we produced the individual SERS cross sections of the 3
modes shown in Fig. \ref{fig2} in Table \ref{tabla}. We believe
these to be the most accurate SERS cross sections ever reported in
the literature to date.

\begin{table}
    \centering
        \begin{tabular}{|c|c|c|c|c|c|}
          \hline
            \multicolumn{6}{|c|}{Rhodamine 6G} \\
          \hline
            Mode  & $\sigma_S$ & $\sigma_{aS}$  & $\tau\sim \hbar/\Gamma$ & $b$ & $T_{eff}$\\
            (cm$^{-1}$) &  (cm$^2$)&  (cm$^2$) &  (ps) &  & (K)\\
            \hline
            610 & 6.0~10$^{-15}$ & 2.2~10$^{-14}$ & 0.8  & 2.6~10$^{-4}$ & 102 \\
            \hline
            1360 & 5.2~10$^{-16}$ & 1.4~10$^{-14}$ & 0.5  & 1.3~10$^{-5}$ & 174 \\
            \hline
            1650 & 1.2~10$^{-15}$ & 9.5~10$^{-14}$ & 0.32  & 2.1~10$^{-5}$ & 220\\
            \hline
        \end{tabular}
    \caption{SERS cross sections for the modes in Fig. \ref{fig2} (676 nm laser) with lifetimes estimated from the FWHM of the peaks. $\sigma_{aS}$ is obtained from $\sigma_{S}$ by means of the asymmetry factor $A$ obtained from the fit. The effective
    mode temperature $T_{eff}$ is obtained from data at 10 K and is mode dependent, thus showing explicitly the breakdown of thermal equilibrium.}
    \label{tabla}
\end{table}

\begin{figure}
\centering{
 \includegraphics[width=8cm]{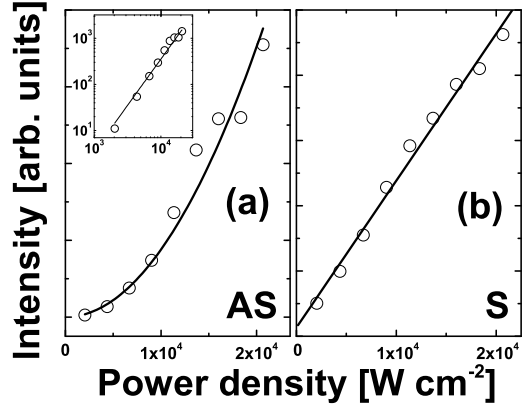}
} \caption{(a) Anti-Stokes signal power dependence at T=20 K with
the corresponding Stokes intensity in (b). From a quadratic fit of
the data in (a) and a linear fit of the data in (b) the SERS cross
section can be obtained. The inset in (a) shows the power
dependence of $I_{aS}$ on a log-log plot, from where a linear fit
with a slope of $\sim 2.1\pm 0.07$ is obtained, thus showing the
expected quadratic behavior. The cross section determined by this
method shows an excellent internal consistency with the
temperature scan in Fig. \ref{fig2}.}
 \label{fig3}
\end{figure}

One interesting consistency check is to reproduce the original
method proposed for SERS pumping\cite{pumping}; i.e. a power
dependence of $I_{S}$ and $I_{aS}$ and compare the values with
those obtained from a temperature scan fitted with Eq.
(\ref{fit}). This is done in Fig. \ref{fig3} for the same laser
line (676 nm, Kr$^{+}$-ion laser) and one mode (610 cm$^{-1}$ of
RH6G). We can stop at a single temperature where the first term in
Eq. (\ref{rho}) dominates (we chose $T=$20 K) and study the
anti-Stokes and Stokes intensities as a function of incident
power. This is shown in Figs. \ref{fig3}(a) and (b). The
anti-Stokes signal is quadratic as a function of incident power
and is fitted with $I_{aS}= \gamma I_L^2$. A plot of $I_{aS}$ on a
log-log graph (inset of Fig. \ref{fig3}(a)) shows indeed that it
has a linear slope of $2.1\pm 0.07$ thus, showing the quadratic
dependence on $I_L$. The Stokes signal is linear in $I_L$ and
fitted to $I_S= \kappa I_L$. From the ratio $\gamma/\kappa=A
\sigma_S \tau / (\hbar \omega_L)$ and a knowledge of the asymmetry
factor $A$ and the lifetime $\tau$, the Stokes cross section
$\sigma_{S}$ can be obtained. The asymmetry factor $A$ can be
obtained from either a measurement at room temperature (where the
second term in Eq. (\ref{rho}) dominates) or by using the $A$
obtained from the temperature scan. From here we obtain
$\sigma_{S}=(7.6 \pm 2.2)~10^{-15}$\,cm$^2$, in excellent
agreement with the temperature scan $\sigma_{S}=(6.0\pm
1.3)~10^{-15}$\,cm$^2$. This shows an internal consistency of the
method and its interpretation, with SERS cross section being
consistent by different approaches. Values obtained at 647 nm (not
shown here) are also slightly larger than those at 676 nm,
consistent with the idea of an underlying resonance\cite{RobJCP}.

We have observed similar pumping effects in most of the important
modes of other classical SERS probes like crystal violet (CV) and
3,3'-diethyloxadicarbocyanine (DODC). Differences in SERS cross
sections for a single analyte (the crossover temperature) can also
be seen by changing the excitation to other lines of the
Kr$^{+}$-laser (647 nm), or lines of the Ar$^{+}$-laser (514 nm);
these results will be reported elsewhere in an extended paper.

We believe our results settle a longstanding debate in the field
of SERS. They also establish a new methodology (different from the
original proposal of power dependence of the anti-Stokes signals
at room temperature) to accurately obtain the SERS cross sections
of different modes. For dyes where the normal cross section
$(\sigma^n)$ can be measured for a given laser, i.e. in situations
where the normal Raman signal of a concentrated solution can be
obtained without interference from fluorescence, this method also
provides an accurate determination of the SERS enhancement factor
for a specific substrate: $G_{{\rm SERS}}\equiv \sigma^{{\rm
SERS}}/\sigma^n$. We do not doubt that there will be many
practical uses of this type of experiments for the determination
of SERS cross sections of important analytes/susbtrates for
applications.

PGE and LFC acknowledge support by EPSRC (UK) under grant
GR/T06124. RCM acknowledges partial support from NPL (UK) and the
hospitality of Victoria University (NZ) where the measurements
have been performed.

\newpage

\end{document}